\documentstyle[aps,pre,eqsecnum,epsf]{revtex}
\newcommand{\be}{\begin{equation}}
\newcommand{\ee}{\end{equation}}
\newcommand{\bea}{\begin{eqnarray}}
\newcommand{\eea}{\end{eqnarray}}

\def \la{\langle}
\def \ra{\rangle}

\begin{document}

\title{Crossover between special and ordinary transitions in 
random semi-infinite Ising-like systems.}

\author{Z. Usatenko$~{^{1,*}}$, Chin-Kun Hu$~{^{2,\$}}$}

\address{\it 
$~{^{1}}$ Institute for Condensed Matter Physics of the National 
Academy of Sciences of Ukraine Lviv, 79011, Ukraine \\ 
$~{^{2}}$ Institute of Physics Academia 
Sinica, Taipei, 11529, Taiwan}

\vspace{0.1cm}
%\date{\today}
\maketitle
\begin{abstract}

We investigate the crossover behavior between special and ordinary 
surface transitions in three-dimensional semi-infinite Ising-like 
systems with random quenched bulk disorder. We calculate the surface 
crossover critical exponent $\Phi$, the critical exponents of 
the layer, $\alpha_{1}$, and local specific heats, $\alpha_{11}$,  
 by applying the field theoretic approach directly in three spatial 
 dimensions ($d=3$) up to the two-loop approximation. The 
 numerical estimates of the resulting 
two-loop series expansions for the surface critical exponents are computed 
by means of Pad\'e and Pad\'e-Borel resummation techniques. We find that 
$\Phi$, $\alpha_{1}$, $\alpha_{11}$  obtained in the
present paper are different from their counterparts of pure Ising
systems. The obtained results confirm that in a system with random quenched 
bulk disorder the plane boundary is characterized by a new set of critical 
exponents. 
\end{abstract}

\renewcommand{\theequation}{\arabic{section}.\arabic{equation}}
\section{Introduction}
\setcounter{equation}{0}

In recent decades the remarkable progress in understanding of the 
critical behavior of real physical systems was achieved from  
application of the powerful field theoretical methods and 
renormalization group (RG) approach to the analysis of these systems. 
It was allowed to perform with higher accuracy the numerical 
analysis of critical exponents and universal amplitude combinations for 
bulk phase transitions \cite{notice1}.  Moreover, these methods have given 
the possibility to investigate the preasymptotic behavior of infinite 
systems. A series of field-theoretical methods developed and tested in the 
studies of bulk phase transitions, have been extended to study the critical 
behavior of systems with boundaries. General reviews on surface critical 
phenomena are given  in Refs. \cite{B83,D86,D97}.

The presence of surfaces, which are 
inevitable in real systems, leads to additional complications. A typical 
model to study the critical phenomena in real physical systems restricted by 
a single planar surface is the semi-infinite model \cite{B83}. As it is 
known from field-theoretic analyses of continuum $\phi^4$ 
model \cite{D86}, the influence of the surface is possible to take into 
account by a quadratic surface term with coeficient $c_{0}$, which describes 
the enhancement of the interactions at the surface, and additional surface 
fields $h_{1}$. There are different surface universality classes, defining the critical 
behavior in the vicinity of boundaries, at temperatures close to the bulk 
critical point ($\tau=(T-T_{c})/T_{c}\to 0$). Each bulk universality class, 
in general, divides into several distinct surface universality classes. 
Three surface universality classes, called ordinary ($c_{0}\to \infty$), 
special ($c_{0}=c_{sp}^{*}$) and extraordinary ($c_{0}\to -\infty$), are 
relevant for our case \cite{D86,D97,DD81}. They meet at a multicritical 
point $(m_{0}^{2},c_{0})=(m_{0c}^{2},c_{sp}^{*})$, which  corresponds 
to the special transition and is called the special point \cite{notice2}.

Most theoretical studies usually concentrate their attention on the 
investigation of critical behaviour distinctly at the fixed points 
$c_{0}=\pm \infty$ and $c_{0}=c_{sp}^{*}$, respectively.
At the present time is very well developed theory of 
critical behavior of individual surface universality classes for pure 
isotropic systems \cite{DD81,DD83,DDE83,DC91,D97,DSh98} and systems with 
quenched surface-enhancement disorder \cite{DN90,PS98,D98}. General 
irrelevance-relevance criteria of the Harris type for 
the systems with quenched short-range correlated surface-bond disorder were  
predicted in \cite{DN90} and confirmed by Monte-Carlo calculations 
\cite{PS98,ILSS98}. Besides, the investigation of the critical behavior of 
the semi-infinite systems with random quenched bulk 
disorder at the ordinary and the special transitions have been studied by us
\cite{UShH00,UH02}. The obtained results 
\cite{UShH00,UH02} have shown that such  
systems are characterized by the new set 
of surface critical exponents in comparison with the case of pure systems.

As it is known, experimental systems are typically characterized by the 
parameters different from the fixed point values. But, the understanding of 
the situation in the crossover regions between different transitions is 
less complete. Investigations of  the crossover behavior between different 
surface universality classes for pure isotropic systems have been published 
in a series of  papers \cite{BL83,DC91,RCz96,CR97,DSh98,MCD99}. However, at 
the present time is open the question about the picture in the crossover 
regions between the different transitions for the semi-infinite systems with 
random quenched bulk disorder. In the present paper we restrict our 
attention to the simplest case with $h_{1}=0$ and investigate crossover 
behavior between special and ordinary transitions for semi-infinite 
Ising-like systems with random quenched bulk disorder. Here it should be 
mentioned that from the whole class of $O(N)$ symmetric $N$-vector models in 
$d$ dimensions only Ising model is the one of primary interest, because it 
satisfies Harris criterion for the specific heat exponent $\alpha(d)\ge 0$ 
\cite{Harris74}. 

The proposed calculations are very important because they 
allow to  understand the phenomenon of adsorption of fluid mixtures in 
contact with a wall, as well as the critical behavior of so-called dilute 
magnets with the surfaces, which can be prepared by mixing an 
(anti)-ferromagnetic material with nonmagnetic one.

The calculations are performed by 
applying field theoretic approach directly in $d=3$ dimentions up to the 
two-loop order approximation. The numerical estimates of the resulting 
two-loop series expansions for the surface crossover exponent $\Phi$ from 
the special to the ordinary transition and surface critical exponents of the 
layer, $\alpha_{1}$, and local specific heats, $\alpha_{11}$, are 
computed by means of the Pad\'e \cite{B75} and Pad\'e-Borel \cite{BNGM76} 
resummation techniques. 
We find that 
$\Phi$, $\alpha_{1}$, $\alpha_{11}$  obtained in the
present paper are different from their counterparts of pure Ising
systems.

\renewcommand{\theequation}{\arabic{section}.\arabic{equation}}
\section{Model}
\setcounter{equation}{0}

The Hamiltonian of the semi-infinite model under consideration 
with random quenched bulk disorder is given by 
\be
H=-\frac{1}{2}\sum_{<i,j>\in bulk}J_{ij}p_{i}p_{j}s_{i}s_{j}-\sum_{<i,j>\in 
surface} J^{'}_{ij}s_{i}s_{j},\label{h1}
\ee
where $s_{i}$ and $s_{j}$ are classical $m$-component spins 
located at the lattice sites $i$ and $j$; a nerest-neighbor bond $<i,j>$ is 
said to belong to the surface region if both $i\in surface$ and $j\in 
surface$, in other cases they belong to the bulk region.
The bulk interaction potential $J_{ij}$ have the parallel to the 
plane translational invariance in the underlying lattice.
 The surface interaction potential $J^{'}_{ij}$ will never be invariant 
with respect to lattice translations parallel to the plane or perpendicular 
to it. The random site variable $p_{i}$ and $p_{j}$ have the probability 
distribution 

$$
P(p_{i})=p\delta(p_{i}-1)+p^{'}\delta(p_{i}),
$$
where $p^{'}=1-p$ is the concentration of nonmagnetic impurities.
 As it is known, there 
are two possible ways to analyze the above random model. The first way is 
connected with direct averaging over random disorder using the method 
introduced by Lubensky \cite{L75}. The second possibility is to perform the 
replica trick $n\to 0$, as it was first done in the renormalization group 
(RG) calculations by Grinstein and Luther \cite{GL76}. Performing 
calculation in the spirit of the method introduced by Grinstein and Luther 
it is possible to show that random model (\ref{h1}) is thermodynamically 
equvalent to the $n$ - vector cubic anisotropic model with effective 
Hamiltonian of the Landau-Ginzburg-Wilson (LGW) type in  
semi-infinite space at the replica limit $n\to 0$ 
\begin{eqnarray} 
H(\vec{\phi}) & = &\int_{0}^\infty dz \int d^{d-1}r [\frac{1}{2} 
\mid \nabla\vec{\phi} \mid ^{2} +  
\frac{1}{2} m_{0}^{2}\mid \vec{\phi} \mid^{2}\nonumber\\
&+& \frac{1}{4!} v_{0} \sum_{i=1}^{n} \phi_{i}^{4} + \frac{1}{4!}u_{0} 
(\mid \vec{\phi}\mid^{2})^{2})] + \int d^{d-1}r  
\frac{1}{2} c_{0}\vec{\phi}^{2},\label{2} 
\end{eqnarray}  
where $\vec{\phi}(x)$ is an $n$-vector field with the 
components $\phi_{i}(x)$, $i=1,...,n$.
Here $m_{0}^2$ is the "bare mass" representing linear measure of the 
temperature difference from the critical point value. The values $u_{0}$ and 
$v_{0}$ are the usual "bare" coupling constants $u_0 < 0$ and $v_{0} > 0$. 
The constant $c_{0}$ is relates to the surface enhancement, which measures 
the enhancement of the interactions at the surface. It should be mentioned 
that the $d$-dimensional spatial integration is extended over a half-space 
$I\!\!R^d_+\equiv\{{\bf x}{=}({\bf r},z)\in I\!\!R^d\mid {\bf r}\in 
I\!\!R^{d-1}, z\ge 0\}$ bounded by a plane free surface at $z=0$. 
The fields $\phi_{i}({\bf r},z)$ 
satisfy the Dirichlet boundary 
condition in the case of ordinary transition: $\phi_{i}({\bf r},z)=0$ at 
$z=0$ and the Neumann boundary condition in the case of 
special transition:    
$\partial_{z}\phi_{i}({\bf r}, z)=0$ at $z=0$ 
\cite{DD81,DDE83}. The model defined in (\ref{2}) is restricted to 
translations parallel to the boundaring surface, $z=0$. Thus, only 
parallel Fourier transformations in $d-1$ dimensions take place.
It should be mentioned that LGW model works good for sufficiently low spin 
dilution $1-p$ as long as system is not too close to the percolation limit.

In order to investigate the critical behavior in the crossover 
region and to calculate the crossover exponent $\Phi$ we should consider 
correlation functions with insertions of the surface operator $\phi_{s}^2$

\be
G^{(N,M;L_{1})}(\{{\bf x}_{i}\},\{{\bf r}_{j}\},\{{\bf R}_{l}\})=\la 
\prod_{i=1}^{N} \phi({\bf x}_{i})\prod_{j=1}^{M}\phi_{s}({\bf r}_{j}) 
\prod_{l=1}^{L_{1}}\frac{1}{2}\phi_{s}^{2}({\bf R}_{l})\ra, \label{7} 
\ee
which involve $N$ fields $\phi({\bf{x}}_{i})$ at distinct points 
${\bf{x}}_{i}(1\leq i \leq N)$ off the surface, $M$ fields 
$\phi({\bf{r}}_{j},z=0)\equiv \phi_{s}({\bf{r}}_{j})$ at distinct surface 
points with parallel coordinates ${\bf{r}}_{j}(1\leq j \leq M)$, and 
$L_{1}$ insertions of the surface operator  
$\frac{1}{2}\vec{\phi}_{s}^{2}({\bf R}_{l})$ ($1 \leq l \leq L_{1}$). 

The corresponding parallel 
Fourier transform of the full free propagator takes form 
\be
G({{\bf{p}}},z,z') = \frac{1}{2\kappa_{0}} \left[ e^{-\kappa_{0}|z-z'|} - 
\frac{c_{0}-\kappa_{0}}{c_{0}+\kappa_{0}} e^{-\kappa_{0}(z+z')} 
\right],\label{8} 
\ee
with the standard notation 
$
\kappa_{0}=\sqrt{p^{2}+m_{0}^{2}}.
$
Here, ${\bf p}$ is the value of parallel momentum associated with $d-1$ 
translationally invariant directions in the system.

\renewcommand{\theequation}{\arabic{section}.\arabic{equation}}
\section{Renormalization}
\setcounter{equation}{0}

The formulation of the renormalization process for the random systems 
introduced by Grinstein and Luther \cite{GL76} is 
essentially the same as in the `pure' case \cite{D86,DSh98}. From other 
side, as it is known from the theory of semi-infinite systems 
\cite{D86,DD81,DD83,DSh98}, the bulk 
field $\phi({\bf x})$ and the surface field $\phi_{s}({\bf r})$ should be 
reparameterized by different uv-finite renormalization factors 
\cite{D86,DSh98} $Z_{\phi}(u,v)$ and $Z_{1}(u,v)$. Thus we have $ \phi = 
Z_{\phi}^{1/2}\phi_{R}$ and  
 $\phi_{s}=Z_{\phi}^{1/2}Z_{1}^{1/2}\phi_{s,R}. $
Besides, introducing the additional surface operator insertions 
$\frac{1}{2}\vec{\phi}_{s}^{2}({\bf R}_{l})$ requires additional specific 
renormalization factor $Z_{\phi_{s}^2}$ 
$$\phi_{s}^{2}=[Z_{\phi_{s}^2}]^{-1}\phi^{2}_{s,R}.$$

The corresponding renormalized correlation functions involving N bulk, M 
surface fields and $L_{1}$ surface operators 
$\frac{1}{2}\vec{\phi}_{s}^{2}({\bf R}_{l})$ can be written as 

\be
G_{R}^{(N,M,L_{1})} ({\bf{p}} ; m,u,v,c)=Z_{\phi}^{-(N+M)/2} Z_{1}^{-M/2} 
Z_{\phi_{s}^{2}}^{L_{1}} G^{(N,M,L_{1})} ({\bf{p}} ; 
m_{0},u_{0},v_{0},c_{0}).\label{9} 
\ee

In the present paper we concentrate our attention on correlation function 
$G^{(0,2,1)} ({\bf{p}} ; m,u,v,c)$ involving two surface fields and 
a single surface operator insertion $\vec{\phi}_{s}^{2}({\bf 
R}_{l})$.

It is well known \cite{DSh98} that the uv-singularities of the correlation  
function $G^{(N,M,L_{1})}$ can be adsorbed 
through a mass shift $m_{0}^{2}=m^2+\delta m^{2}$ and surface-enhancement 
shift $c_{0}=c+\delta c$. The renormalizations of the mass $m$, 
coupling constant $u, v$ and the renormalization factor $Z_{\phi}$ are 
defined by standart normalization conditions of the infinite-volume theory 
\cite{BGZ76,GL76,P80,PS00,PV00}. In order to adsorb uv singularities located 
in the vicinity of the surface, a surface-enhancement shift $\delta c$ 
is required. In this connection the new normalization condition should be 
introduced (see Appendix 1). Taking into account the normalization condition 
(\ref{12})  and expression for renormalized correlation function (\ref{9}) 
it is possible to define the renormalization factor $Z_{\phi^2}$ in the 
form 
\be
[Z_{\phi_{s}^2}]^{-1}=\left. Z_{\parallel}\frac{\partial 
[G^{(0,2)}(0;m_{0},u_{0},v_{0},c_{0})]^{-1}}{\partial 
c_{0}}\right|_{c_{0}=c_{0}(c,m,u,v)}.\label{14} 
\ee

It should be mentioned that renormalization factor $Z_{\parallel} = Z_{1} 
Z_{\phi}$ is defined via the standard normalization condition (\ref{11}) 
(see \cite{DSh98},\cite{UH02}) 
 
\be
Z_{\parallel}^{-1} = \left. 2m \frac{\partial}{\partial 
p^{2}}[G^{(0,2)} (p)]^{-1}\right|_ {p^2=0} = \lim_{p\to 
0}{m\over p}{\partial\over\partial p} [G^{(0,2)}(p)]^{-1}.\label{15} 
\ee

Eq. (\ref{14}) enables us 
considerably simplify the calculation of the correlation function 
$G^{(0,2,1)}$ with surface operator $\vec{\phi}_{s}^{2}({\bf 
R}_{l})$ insertion.

It should be noted that all $Z$ factors in the $d<4$ case  
have finite limits at $\Lambda \to \infty$ (where $\Lambda$ is a 
large-momentum cutoff). All factors mentioned above depend on 
the dimensionless variables $u$ and $v$. Besides, the surface 
renormalization factors $Z_{1}$ and $Z_{\phi_{s}^2}$ depend on both 
$u,v$ and the dimensionless ratio $c/m$. The last dependence on the ratio 
$c/m$ plays the crucial role in the investigation of the crossover behavior 
from the special transition ($c/m\to 0$) to the ordinary transition ($c/m\to 
\infty$). 

\renewcommand{\theequation}{\arabic{section}.\arabic{equation}} 
\section{Expansion of the correlation function near the multicritical point} 
\setcounter{equation}{0}

As was indicated before, the main goal of the present work is to investigate 
the scaling critical behavior between special and ordinary transition and 
to calculate the crossover exponent $\Phi$. In this connection let 
consider the small deviations $\Delta c_{0}=c_{0}-c_{sp}^{*}$ from the 
multicritical point. The power expansion of the bare correlation 
functions $G^{(N,M)}({\bf{p}};m_{0},u_{0},v_{0},c_{0})$ in terms of  this 
small deviations $\Delta c_{0}$ has a form \be 
G^{(N,M)}({\bf{p}};m_{0},u_{0},v_{0},c_{0})=\sum_{L_{1}=0}^{\infty} 
\frac{(\Delta c_{0})^{L_{1}}}{L_{1} !} 
G^{(N,M,L_{1})}({\bf{p}};m_{0},u_{0},v_{0},c_{sp}^{*}).\label{16} \ee

Based on Eq.(\ref{9}), we rewrite the right-hand part of Eq.(\ref{16}) in 
terms of the renormalized correlation functions 
 and  renormalized variable 
 $ \Delta c=[Z_{\phi^{2}_{s}}(u,v)]^{-1} \Delta c_{0} $
and obtain 
\begin{eqnarray}
&& Z_{\phi}^{-(N+M)/2} 
(Z_{1})^{-M/2}G^{(N,M)}({\bf{p}};m_{0},u_{0},v_{0},c_{0})\nonumber\\
&& = \sum_{L_{1}=0}^{\infty}\frac{(\Delta c)^{L_{1}}}{L_{1} !} 
G^{(N,M,L_{1})}_{R}({\bf{p}};m,u,v).\label{18} \end{eqnarray}

The last equation in straightforward fashion define the correspondent 
renormalized correlation functions defined in the vicinity of the 
multicritical point 

\be
G^{(N,M)}_{R}({\bf{p}};m,u,v,\Delta c)=Z_{\phi}^{-(N+M)/2} 
(Z_{1})^{-M/2}G^{(N,M)}({\bf{p}};m_{0},u_{0},v_{0},c_{0}).\label{19}
\ee
It is easy to see that these correlation functions depend on the 
dimensionless variable $ \bar{c}=\Delta c/m.$
Thus, the correlation functions 
$G^{(N,M)}_{R}({\bf{p}};m,u,v,\Delta c)$ satisfy correspondent 
Callan-Symanzik equations \cite{Sh97,DSh98} 
\begin{eqnarray}
&& \left[ m\frac{\partial}{\partial m}+\beta_{u} 
(u,v)\frac{\partial}{\partial u}+ \beta_{v} (u,v)\frac{\partial}{\partial 
v}+\frac{N+M}{2}\eta_{\phi}(u,v)\right.\nonumber\\ 
&& +\left.\frac{M}{2}\eta^{sp}_{1}(u,v)-[1+\eta_{\bar{c}}(u,v)] 
\bar{c}\frac{\partial}{\partial \bar{c}}\right] 
G^{(N,M)}_{R}({\bf{p}};m,u,v,\Delta c)= \Delta G_{R},\label{21}
\end{eqnarray}
where the inhomogeneous part $\Delta G_{R}$ should be negligible in the 
critical region similarly as that takes place in the case of infinite field 
 theory. The resulting homogeneous equation differs from the standart bulk 
Calan-Symanzik (CS) equation \cite{Z89,Parisi88,ID89} in that fashion it 
involves the additional surface related function $\eta^{sp}_{1}$ and 
term $-[1+\eta_{\bar{c}}(u,v)] \bar{c}\frac{\partial}{\partial 
\bar{c}}$, where 
\be
\eta_{1}^{sp}(u,v)=\left. m\frac{\partial}{\partial m}\right|_{FP} ln 
Z_{1}(u,v)=\left.\beta_{u}(u,v)\frac{\partial ln Z_{1}(u,v)}{\partial 
u}+ \beta_{v}(u,v)\frac{\partial ln Z_{1}(u,v)}{\partial 
v}\right|_{FP}\label{22} \ee and
\be
\eta_{\bar{c}}(u,v)=\left. m\frac{\partial}{\partial m}\right|_{FP} ln 
Z_{\phi_{s}^2}(u,v)=\left.\beta_{u}(u,v)\frac{\partial ln 
Z_{\phi_{s}^2}(u,v)}{\partial u}+ \beta_{v}(u,v)\frac{\partial ln 
Z_{\phi_{s}^2}(u,v)}{\partial v}\right|_{FP}.\label{23} \ee

It should be mentioned that functions $\beta_{u}(u,v)$, 
$\beta_{v}(u,v)$ and $\eta_{\phi}(u,v)$ appearing in (\ref{21}) are the 
usual bulk RG functions.
The symbol 'FP' indicates that the above value should be calculated at the 
infrared-stable random fixed point (FP) of the underlying bulk theory.

\renewcommand{\theequation}{\arabic{section}.\arabic{equation}} 
\section{Scaling critical behavior at the multicritical point} 
\setcounter{equation}{0}

The asymptotic scaling critical behavior of the correlation functions 
can be obtained through detailed analysis of the CS equations of Eq. 
(\ref{21}), as was proposed in \cite{Z89,BB81} and 
employed in the case of the semi-infinite systems in 
\cite{CR97,DSh98,MCD99}. Our present investigations of the scaling 
critical behavior are in complete analogy with the scheme mentioned above 
\cite{BB81,DSh98} (see Appendix 2). Taking into account the scaling form of 
the renormalization factor $Z_{\phi_{s}^2}$ of Eq. (\ref{25}) and the 
relation $m\sim \tau^{\nu}$, we obtain for $\Delta c$ and for the 
scaling variable $\bar{c}$ the next asymptotic dependences 
\be
\Delta c\sim m^{-\eta_{\bar{c}}(u^{*},v^{*})} \Delta 
c_{0},\quad\quad \Delta c\sim \tau^{-\nu 
\eta_{\bar{c}}(u^{*},v^{*})} \Delta c_{0}\label{26} \ee
and
\be
\bar{c}\sim m^{-(1+\eta_{\bar{c}}(u^{*},v^{*}))} \Delta 
c_{0},\quad\quad\quad \bar{c}\sim \tau^{-\Phi} \Delta c_{0},\label{27}
\ee
where
\be
\Phi=\nu 
(1+\eta_{\bar{c}}(u^{*},v^{*})) \label{28}
\ee
is the surface crossover critical exponent. Eq. (\ref{27}) 
explains the physical meaning of the surface crossover exponent as a value 
which characterises the measure of deviation from the multicritical point.  
The second equations in Eqs.(\ref{26}) and (\ref{27}) indicate about 
non-analitic temperature dependence of the renormalized surface-enhancement 
deviation $\Delta c$. 
Taking into account the above mentioned results from the CS equation we 
obtain the next asymptotic scaling form of the surface correlation function 
$G^{(0,2)}$

\begin{eqnarray}
&& G^{(0,2)}(p;m_{0},u_{0},v_{0},c_{0})\sim 
m^{-\frac{\gamma_{11}^{sp}}{\nu}} 
G^{(0,2)}_{R}(\frac{p}{m};1,u^{*},v^{*},m^{-\Phi / \nu}\Delta 
c_{0})\nonumber\\ && \sim 
\tau^{-\gamma_{11}^{sp}}G(pt^{-\nu};1,\tau^{-\Phi}\Delta c_{0}), \label{29} 
\end{eqnarray} where
$
\gamma_{11}^{sp}=\nu (1-\eta_{\parallel}), 
$
is the local surface susceptebility exponent and 
$
\eta_{\parallel}^{sp}=\eta_{1}^{sp}+\eta 
$
is the surface correlation exponent \cite{mit1}.
It is easy to see, that the asymptotic scaling critical behavior of the 
surface correlation function for the systems with random  
quenched bulk disorder is characterized by the new crossover exponent 
$\Phi(u^{*},v^{*})$, which belongs to the universality class of the random 
model. In the next section, we will calculate the surface crossover 
exponent $\Phi$ of the semi-infinite systems with random quenched bulk 
disorder. 

\renewcommand{\theequation}{\arabic{section}.\arabic{equation}}
\section{The perturbation series up to two-loops.} 
\setcounter{equation}{0}

According to the Eqs.(\ref{28}) and (\ref{23}) the calculation 
of the crossover critical exponent $\Phi$ is connected with calculation 
of the renormalization factor $Z_{\phi_{s}^2}$ via Eq.(\ref{14}). 
The usual bulk {\it uv}-singularities which are present in correlation
function $[G^{(0,2)}(0)]^{-1}$ can be removed by the method similar 
to those repurted in Refs.\cite{P80,DSh94,DSh98,UShH00} with help of 
standard mass renormalization procedure. 

The second step of our calculation is to remove the uv divergences which are 
connected with the presence of the surface in the system. The 
surface uv-singularities of the inverse surface correlation function  
$[G^{(0,2)}(0)]^{-1}$ can be removed by performing the surface enhancement 
renormalization which is defined by Eq.(\ref{10}). For convenience we can 
rewrite the normalization condition of Eq. (\ref{10}) in the form 
\be
Z_{\parallel}[G^{(0,2)}(0;m_0,u_0,v_0,c_0)]^{-1}=m + c.\label{34}
\ee
for inverse unrenormalized surface correlation function 
$[G^{(0,2)}(0)]^{-1}$. Performing the differentiation 
of the above mentioned normalization condition with respect to 
$\frac{\partial}{\partial c_{0}}$ and taking into account Eq.(\ref{14}) we 
obtain for the renormalization factor $Z_{\phi^2}$ the next equation 
\be
Z_{\phi_{s}^2}=\frac{\partial c_{0}}{\partial c},\label{35}
\ee
where $c_{0}=c+\delta c$ and 
\be
\delta c = (Z_{\parallel}^{-1}-1)(m+c) + \sigma_{0}(0;m,c_0=c+\delta 
c).\label{36}
\ee
Here $\sigma_{0}(0;m,c_0)$ denotes the sum of loop diagrams of all orders in 
$[G^{(0,2)}(0;m,u_{0},v_{0},c_{0})]^{-1}$ (see \cite{DSh98,UH02}). 
Among them $\sigma_{1}$ corresponds to the one-loop graph, $\sigma_{2}$ 
denotes the melon-like two-loop diagrams 
\be
\sigma_{2}= \raisebox{-5pt}{\epsfxsize=1.7cm \epsfbox{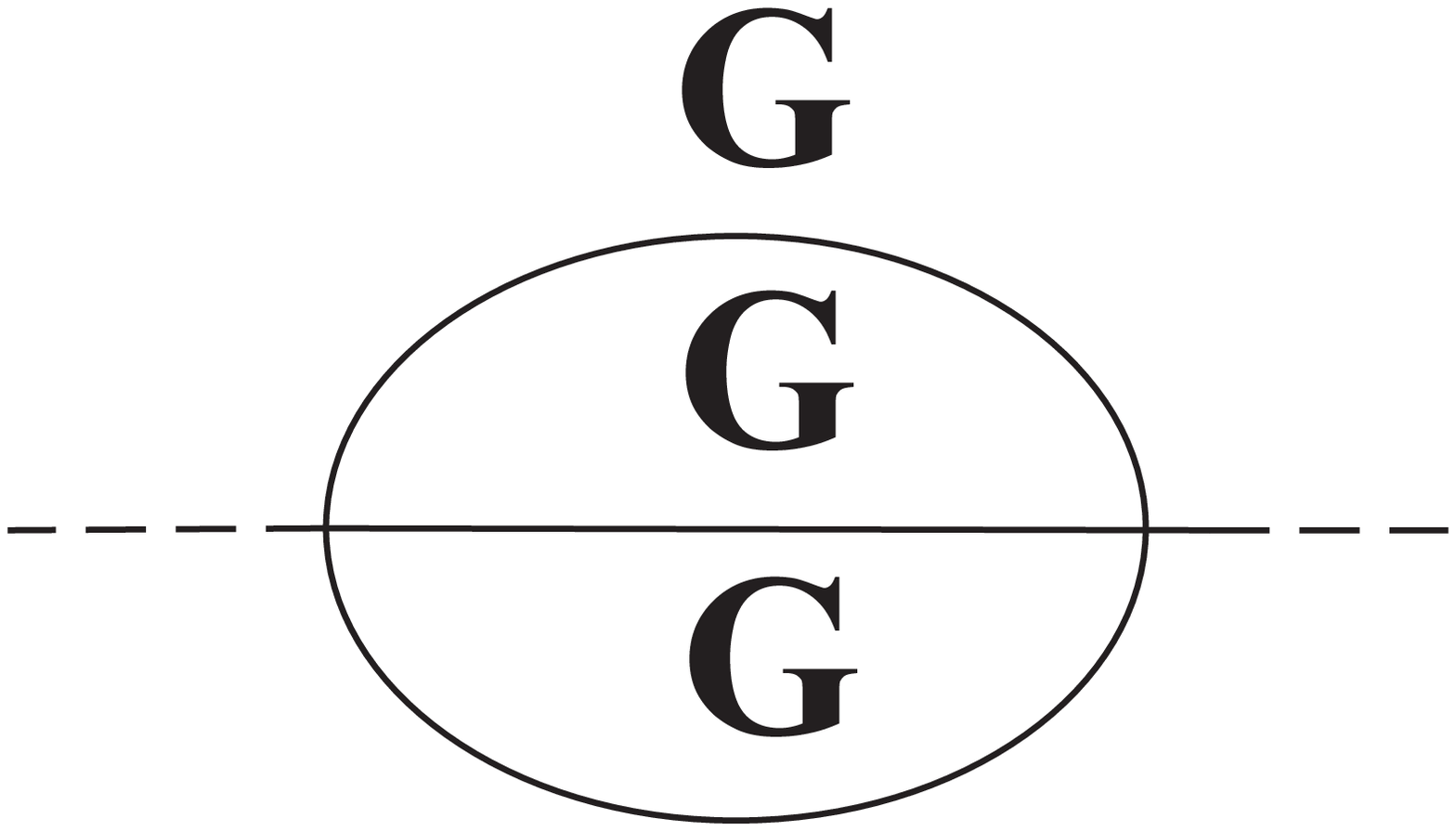}}
-\frac{1}{2\kappa}\raisebox{-5pt}{\epsfxsize=0.9cm \epsfbox{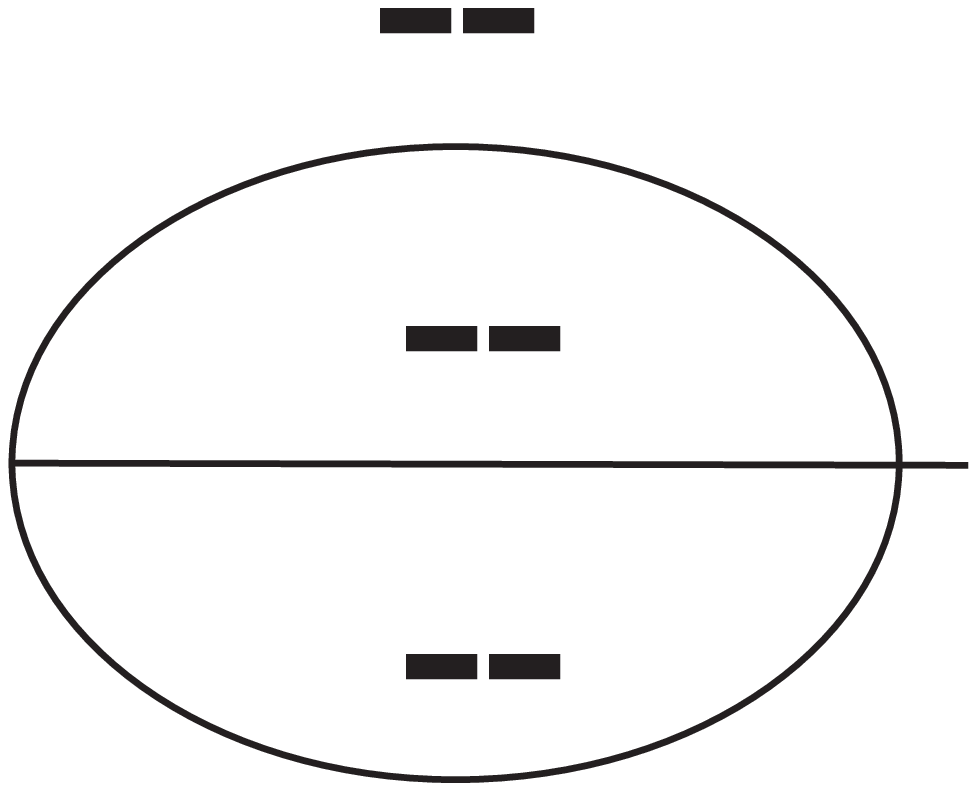}} +
\frac{m^2}{2\kappa}\frac{\partial}{\partial k^2}\left. 
\raisebox{-5pt}{\epsfxsize=1cm 
\epsfbox{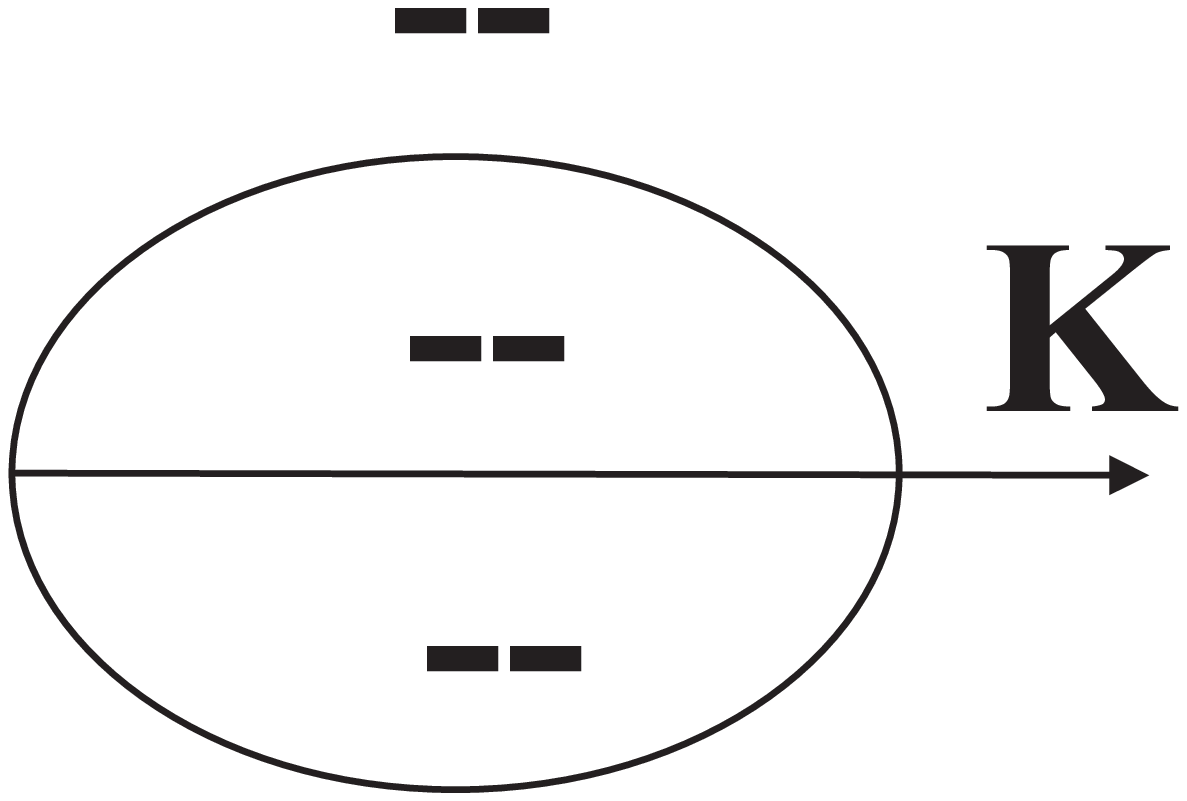}}\right|_{k^2=0},\label{37}
\ee

$\sigma_{3}$ and $\sigma_{4}$ represent the reducible and irreducible 
two-loop diagrams in $[G^{(0,2)}(0;m,u_{0},v_{0},c_{0})]^{-1}$, 
respectively. Here the full lines with labels "G" denote the full free 
propagator of Eq.(\ref{8}). Eq. (\ref{35}) can be 
resolved by using the method of sequential iteration. As a result of the 
first order of the perturbation theory (one-loop approximation) at general 
spatial dimensions $d<4$, we obtain 
\be 
Z_{\phi_{s}^2}^{(1)}=1-\frac{T_{1}}{2}\frac{\pi^{\frac{-(d-1)}{2}}}{16} 
\frac{\Gamma(\epsilon)}{\Gamma(\frac{\epsilon+1}{2})}[1-2^{\frac{1+\epsilon}{2}} 
{}_{2}F_{1}(\frac{3-\epsilon}{2};\frac{1+\epsilon}{2};
\frac{3+\epsilon}{2};\frac{1}{2})],\label{38} \ee 
where ${}_{2}F_{1}(...)$ is the hypergeometric function and 
coefficient $T_{1}=\frac{n+2}{3}\bar{u_{0}}+\bar v_{0}$ includes 
dimensionless effective expansion parameters $\bar u_{0}=u_{0}m^{-\epsilon}$ 
and $\bar v_{0}=v_{0}m^{-\epsilon}$ which are identified through the 
standart vertex renormalization. In the case of massive theory the value of 
$\epsilon$ is not necessary restricted to be a small and the above 
expressions hold for any relevant dimensions $d<4$. The presence of the 
Euler gamma function $\Gamma(\epsilon)$ indicates the existence of 
dimensional poles $1/\epsilon$ when $\epsilon\to 0$. 

According to Eqs.(\ref{14}) 
and (\ref{38}) for $\eta_{\bar{c}}$ at a one-loop 
order in the case of general dimensions up to $d=4$ we obtain 
\be
\eta_{\bar{c}}=\frac{T_{1}}{2}\frac{\epsilon}{1+\epsilon}\pi^{-d/2}2^{-d}\Gamma(\epsilon/2)[1-2^{\frac{1+\epsilon}{2}}
{}_{2}F_{1}(\frac{3-\epsilon}{2};\frac{1+\epsilon}{2};\frac{3+\epsilon}{2};
\frac{1}{2})].\label{39}
\ee

At the random fixed point of order 
$O(\sqrt{\epsilon})$ \cite{OO92}($K_{4}u^{*}=-3\sqrt{\frac{6\epsilon}{53}},
K_{4}v^{*}=4\sqrt{\frac{6\epsilon}{53}}$, where the geometric 
normalization factor $K_{4}=1/(8\pi^2)$ ), Eq.(\ref{39}) in the limit 
$\epsilon\to 0$ leads to

\be
\lim_{\epsilon\to 
0}\eta_{\parallel}^{sp}=\lim_{\epsilon 
\to 0}\eta_{\bar{c}}=-\sqrt{\frac{6\epsilon}{53}}.\label{40} \ee 
This result coincides with that obtained by Ohno and Okabe\cite{OO92}.
At $\epsilon\to 0$ for the surface crossover exponent $\Phi$ 
in one-loop calculations we obtain 
\be
\Phi=\frac{1}{2}-\frac{1}{4}\sqrt{\frac{6\epsilon}{53}}.\label{41}
\ee
In the case of three spatial dimensions ($d=3$) the renormalization 
factors $Z_{1}$ and $Z_{\phi_{s}^2}$ are finite and their one-loop 
expressions do not coincide. At one-loop order, we 
obtain 
\be \eta_{\bar{c}}\approx -0.596 \quad\quad and 
\quad\quad \Phi=0.286.\label{42} 
\ee
In the next order of the perturbation theory we restrict our attention only 
to the case of $d=3$ dimensions. 
Thus after the surface-enhancement renormalization and performing the 
Feynman integrals in analogy with \cite{DSh98,UH02} and carrying out  
the vertex renormalizations of bare dimensionless 
parameters  $\bar{u}_{0}=u_{0}/8\pi m$ and $\bar{v}_{0}=v_{0}/8\pi m$
\begin{eqnarray}
&& \bar{u}_{0}=\bar{u}(1+\frac{n+8}{6}\bar{u}+\bar{v}),\nonumber\\
&& \bar{v}_{0}=\bar{v}(1+\frac{3}{2}\bar{v}+2\bar{u}),\label{43} 
\end{eqnarray}
we obtain a second-order series expansion for the renormalization factor 
$Z_{\phi_{s}^2}$ in terms of new renormalized coupling constants $\bar{u}$ 
and $\bar{v}$, 
\be
Z_{\phi_{s}^2}(\bar{u},\bar{v})=1+\frac{n+2}{3}(ln2-\frac{1}{4})\bar{u}+
(ln2-\frac{1}{4})\bar{v}+\frac{n+2}{3}C(n)\bar{u}^2+2C(n)\bar{u}\bar{v}+C(1)\bar{v}^2,\label{44}
\ee
where $C(n)$ is a function of the replica number $n$, defined by
\be
C(n)=A-B-\frac{n}{2}ln2+\frac{n+2}{2}ln^{2}2+\frac{2n+1}{12},\label{45}
\ee

and $A=0.202428$, $B=0.678061$ are 
integrals originating from the two-loop melon-like diagrams. 
Combining the renormalization 
factor $Z_{\phi_{s}^2}$ with the one-loop pieces of the $\beta$ functions 
$\beta_{\bar{u}}(\bar{u},\bar{v})=-\bar{u}(1-[(n+8)/6]\bar{u}-\bar{v})$ and 
$\beta_{\bar{v}}(\bar{u},\bar{v})=-\bar{v}(1-\frac{3}{2}\bar{v}-2\bar{u})$ 
according to Eg.(\ref{23}), we obtain the desired series expansion for 
$\eta_{\bar{c}}$, 
\be
\eta_{\bar{c}}(u,v)=-2\frac{n+2}{n+8}(ln2-\frac{1}{4})u-\frac{2}{3}(ln2-\frac{1}{4})v-
8[3\frac{n+2}{(n+8)^2}D(n)u^2+\frac{2D(n)}{n+8}uv+\frac{D(1)}{9}v^2],\label{46}
\ee 
where 
\be
D(n)=A-B+\frac{n+2}{3}ln^{2}2-\frac{n+1}{2}ln2+\frac{17n+22}{96},\label{47}
\ee
and renormalized coupling constants $u$ and $v$, normalized in a standard 
fashion $u=[(n+8)/6]\bar{u}$ and $v=\frac{3}{2}\bar{v}$. In common, 
Eq.(\ref{46}) gives a result for the model with the effective Hamiltonian 
of the Landau-Ginzburg-Wilson type with cubic anisotropy in the 
semi-infinite space (\ref{2}) with general number $n$ of order parameter 
components. Our calculations are connected with the 
investigation of the critical behavior of {\it{semi-infinite random 
Ising-like}} systems by taking the replica limit $n\to 0$. Hence, we 
obtain 
\be
\eta_{\bar{c}}=-\frac{1}{2}(ln2-\frac{1}{4})u-\frac{2}{3}(ln2-\frac{1}{4})v-
8[\frac{3}{32}D(0)u^2+\frac{D(0)}{4}u v+\frac{D(1)}{9}v^2].\label{48}
\ee
The knowledge of $\eta_{\bar{c}}$ gives access to the calculation of 
the crossover critical exponents $\Phi$ via the scaling relation 
of Eq.(\ref{28}). Besides, we can calculate the critical exponents 
$\alpha_{1}$ and $\alpha_{11}$ of the layer and specific heats via the usual 
scaling relations \cite{D86}
\be
\alpha_{1}=\alpha+\nu-1+\Phi=1-\nu (d-2-\eta_{\bar{c}}),\quad\quad 
\alpha_{11}=\alpha+\nu-2+2\Phi = -\nu 
[d-3-2\eta_{\bar{c}}].\label{49} 
\ee

Above critical exponents should be calculated at the standard 
infrared-stable random  fixed (FP) point of the underlying bulk theory 
\cite{Jug83} $u^{*}=-0.60509$ and $v^{*}=2.39631$, as it is usually accepted 
in the massive field theory. 

\section{Numerical results}

For each of the surface critical exponents mentioned above and 
crossover exponent $\Phi$ we obtain from Eq.(\ref{48}) at $d=3$ a double 
series expansion in powers of $u$ and $v$ truncated at the second order 
\cite{mit2}. In order to perform the analysis of these perturbative series 
expansions and to obtain accurate estimates of the surface critical 
exponents a powerful resummation procedure must be used. One of the simplest 
ways is to perform the double Pad\'e-analysis \cite{B75}. This 
should work well when the series behave in lowest orders "in a convergent 
fashion". Another way is to perform the double Pad\'e-Borel 
analysis \cite{BNGM76} for these series. The usage of the 
Pad\'e-Borel resummation procedures are posible in the case when the 
terms in the series are alternating in sign \cite{BNM78}. The results of our 
calculations at Jug random fixed point \cite{Jug83} are represented in 
Table 1. The quantities $O_{1}/O_{2}$ and $O_{1i}/O_{2i}$ represent the 
ratios of magnitudes of first-order and second-order perturbative 
corrections appearing in direct and inverse series expansions. The larger 
(absolute) value of these ratios indicate about the better apparent 
convergence of truncated series.

The values $[p/q]$ (where $p,q=0,1$) in Table 1 are 
simply Pad\'e approximants which represent the partial sums of the direct 
and inverse series expansions up to the first and second order. The 
nearly diagonal two-variable rational approximants of the types $[11/1]$ and 
$[1/11]$ give at $u=0$ or $v=0$ the usual 
$[1/1]$ Pad\'e approximant \cite{B75}. The results of Pad\'e-Borel analysis 
of the direct $R$ and the inverse $R^{-1}$ series expansions give numerical 
estimates of the surface critical exponents with a high degree of 
reliability. As it is easy to see, the most reliable estimate is obtained 
from the inverse series expansion for the surface critical exponent 
$\alpha_{1}$, which represent the best convergence properties. 
Substituting this value $\alpha_{1}=0.211$ together with the standard bulk 
value $\nu=0.678$ into the scaling relations (\ref{28}) and (\ref{49}), we 
have obtained the remaining critical exponents that are present in the last 
column of Table 1. The deviations of these estimates from the other 
estimates of the table give a rough measure of the achieved numerical 
accuracy.

The results of the similar analysis of the perturbative series expansions 
of the surface critical exponent at random fixed point $u^{*}=-0.6524, 
v^{*}=2.4203$ \cite{Sh88} are presented in Table 2 for comparison. In a 
similar way the most reliable estimate is obtained for inverse series 
expansion of $\alpha_{1}$. The results of substituting of 
$\alpha_{1}=0.208$ and $\nu=0.679$ into scaling laws 
(\ref{28}) and (\ref{49}) are presented in the last column of Table 2. As 
it easy to see from comparison of the results Table 1 and Table 2, the 
difference in the ways of the $\beta$ functions resummation have not 
essential influence on the values of the surface critical exponents. 
The difference between final results of Table 1 and Table 2 are $1.2\%$ for 
$\eta_{\bar{c}}$, $1.4\%$ for $\alpha_{1}$, $1.8\%$ for $\alpha_{11}$ and 
$0.2\%$ for $\Phi$.

For evaluation of the reliability of the results obtained in the 
two-loop approximation, we have performed additional calculation of the
surface critical exponents from $\alpha_{1}=0.211$ and six-loop perturbation 
theory results \cite{PV00} for bulk critical exponent of the correlation 
length $\nu=0.678(10)$. We have obtained $\eta_{\bar{c}}=-0.164$, 
$\alpha_{11}=-0.222$ and for surface crossover critical exponent 
$\Phi=0.567$. This indicates about good stability of our results obtained in 
the frames of two-loop approximation scheme.

\section{Summary} 

We have studied the crosover critical behavior between special and 
ordinary surface transitions of three-dimensional quenched random 
semi-infinite Ising-like systems. We find that the asymptotic scaling 
critical behavior of the surface correlation function for the systems with 
random  quenched bulk disorder is characterized by the new crossover 
exponent $\Phi(u^{*},v^{*})$, which belong to the universality class of the 
random model. We have calculated the 
crossover critical exponent $\Phi$ and critical exponents of the layer, 
$\alpha_{1}$, and local specific heats exponent, $\alpha_{11}$, by applying 
the field theoretic approach directly in three dimentions up to the two-loop 
approximation. We performed a rational double Pad\'e and double Pad\'e-Borel 
analysis of the resulting perturbation series expansions for the surface 
critical exponents in order to find the best numerical values. The final 
numerical values of the surface critical exponents $\alpha_{1}$, 
$\alpha_{11}$ and crossover exponent $\Phi$ for the systems with quenched 
random bulk disorder are 
\be
\alpha_{1}=0.211\pm 0.003\quad\quad\quad \alpha_{11}=-0.222\pm 
0.004\quad\quad\quad \Phi=0.567\pm 0.001.
\ee

These values evidently different from their counterparts of pure Ising 
systems \cite{DSh94,DSh98}
\be
\alpha_{1}=0.279\quad\quad\quad \alpha_{11}=-0.182\quad\quad\quad \Phi=0.539.
\ee

We suggest that the obtained results could stimulate further experimental 
and numerical investigations of the surface critical behavior of random 
systems.

\section*{Acknowledgments}
Dr.C.-K.Hu would like to thank the National Science Council of Republic of 
China (Taiwan) for supporting of this work under Grant No. NSC 
91-2112-M-001-056.

\renewcommand{\theequation}{A1.\arabic{equation}}
\section*{Appendix 1}
\setcounter{equation}{0}

In order to specify $\delta c$, $Z_{1}$ and $Z_{\phi^2_{s}}$, we require 
that \cite{DSh94,DSh98}
\be
\left.G_{R}^{(0,2)}({\bf{p}};m,u,v,c)\right|_{{\bf{p}}=0} = 
\frac{1}{m+c},\label{10} \ee

\be
\left.\frac{\partial G_{R}^{(0,2)} ({\bf{p}};m,u,v,c)}{\partial p^{2}} 
\right|_{{\bf{p}}=0} = - 
\frac{1}{2m(m+c)^{2}}, \label{11}
\ee
and correspondent normalization condition for the correlation function 
$G^{(0,2,1)}$ with the insertion of the surface operator 
$\frac{1}{2}\phi_{s}^2$ 

\be
\left.G_{R}^{(0,2,1)}({\bf{p}};m,u,v,c)\right|_{{\bf{p}}=0}=\frac{1}{(m+c)^2}.\label{12}
\ee

Eq.(\ref{12}) is motivated by the fact that the bare 
correlation function $G^{(0,2,1)}(0;m_{0},u_{0},v_{0},c_{0})$ may be 
written as derivative 
$-\frac{\partial}{\partial c_{0}}G^{(0,2)}(0;m_{0},u_{0},v_{0},c_{0})$.
This equation simplifies considerably the 
calculation of the correlation function with 
insertions of surface operator $\frac{1}{2}\phi_{s}^2$. 

From Eq.(\ref{10}), it is easy to see that the special point is located at 
$m=c=0$, because at this point the divergence of the bulk and the surface 
correlation length and susceptibility is observed. At $c=0$ the surface 
normalization conditions are simplified and yield $c_{0}=c_{sp}^{*}$. This 
point corresponds to the multicritical point $(m_{0c}^{2},c_{sp}^{*})$ at 
which special transition takes place. On the other hand, the above 
mentioned equation implies also that the surface correlation length and the 
susceptibility are finite at the ordinary transition, because in this case 
we have $c>0$ when $m \to 0$. This latter case corresponds to the situation 
when the surface remains "noncritical" at the bulk transition temperature.

\renewcommand{\theequation}{A2.\arabic{equation}}
\section*{Appendix 2}
\setcounter{equation}{0}

As it is usually accepted in the 
massive field theory, the variable $m$ is identified with 
the inverse bulk correlation length $\xi^{-1}$ and is 
proportional to $\tau^{\nu}$, where $\tau=(T-T_{c})/T_{c}$. Following the 
scheme proposed in \cite{BB81}, we can perform the integration of 
Eq.(\ref{22}), Eq.(\ref{23}) and expressions for the RG functions 
$\beta_{u}(u,v)$, $\beta_{v}(u,v)$ and $\eta_{\phi}(u,v)$. This gives the 
following asymptotic dependencies at $m\to 0$ 
\begin{eqnarray}
&& |u-u^{*}|\sim m^{\omega_{u}},\quad\quad where \quad\quad
\omega_{u}=\beta^{'}_{u}(u^{*},v^{*}),\nonumber\\ 
&& |v-v^{*}|\sim m^{\omega_{v}},\quad\quad where \quad\quad
\omega_{v}=\beta^{'}_{v}(u^{*},v^{*}),\nonumber\\
&& Z_{\phi}\sim m^{\eta},\quad\quad\quad\quad where \quad\quad
\eta=\eta_{\phi}(u^{*},v^{*}),\nonumber\\
&& Z_{1}\sim m^{\eta^{sp}_{1}(u^{*},v^{*})},\nonumber\\
&& Z_{\phi_{s}^{2}}\sim 
m^{\eta_{\bar{c}}(u^{*},v^{*})}.\label{25} \end{eqnarray} 

As follows from these expressions, the variables $u$ and $v$ deviate from 
their fixed values $u^{*}$ and $v^{*}$ by different scaling laws with 
various values of $\omega_{u}$ and $\omega_{v}$. The 
scaling laws (see Eq.(\ref{25})) have the similar form as in the case of the 
pure systems \cite{DSh98}, but renormalization factors are characterized by 
another values of the critical exponents which belong to the universality 
class of random model.

\begin{table}[htb]
\caption{Surface critical exponents involving the RG function 
$\eta_{\bar{c}}$ at the Jug fixed point 
$u^*=-0.60509,v^*=2.39631$ (two-loop order) } \label{tab1} \begin{center}
\begin{tabular}{rrrrrrrrrrrrr}
\hline
$ exp $~&~
$\frac{O_{1}}{O_{2}}$~&~$\frac{O_{1i}}{O_{2i}}$~&~$[0/0]$~&~
$[1/0]$~&~$[0/1]$~&~$[2/0]$~&~$[0/2]$~&~$[11/1]$~&~$[1/11]$~&~$[R]$~&~$R_{i}^{-1}$~&~$f(\alpha_{1},\nu,\eta)$
\\ 
\hline
$\eta_{\bar{c}}$ & -0.8 & -1.5 & 0.00 & -0.574 & -0.365 & 0.150 & -0.152 
 & -0.281 & -0.268 & -0.313 & -0.280 & -0.164\\

$\alpha_{1}$ & -1.7 & -7.6 & 0.50 & 0.051 & 0.190 & 0.312 & 0.220 & 0.201 
& 0.213 & 0.185 & 0.211 & 0.211\\

$\alpha_{11}$ & -1.1 & -2.8 & 0.00 & -0.574 & -0.365 & -0.036 & -0.268 & 
-0.324 & -0.306 & -0.351 & -0.313 & -0.222\\

$\Phi$ & -0.5 & -0.5 & 0.5 & 0.375 & 0.389 & 0.652 & 0.658 & 0.451 & 0.452 
& 0.444 & 0.445 & 0.567\\

\end{tabular}
\end{center}
\end{table}

\begin{table}[htb]
\caption{Surface critical exponents involving the 
RG function $\eta_{\bar{c}}$ at the fixed point 
$u^*=-0.6524,v^*=2.4203$ (two-loop order)}
 
\label{tab2} 
\begin{center} 
\begin{tabular}{rrrrrrrrrrrrr} 
\hline $ exp $~&~ 
$\frac{O_{1}}{O_{2}}$~&~$\frac{O_{1i}}{O_{2i}}$~&~$[0/0]$~&~ 
$[1/0]$~&~$[0/1]$~&~$[2/0]$~&~$[0/2]$~&~$[11/1]$~&~$[1/11]$~&~$[R]$~&~$R_{i}^{-1}$~&~$f(\alpha_{1},\nu,\eta)$ 

\\ \hline $\eta_{\bar{c}}$ & -0.82 & -1.55 & 0.0 & -0.570 & -0.363 & 0.124 & 
-0.168 & -0.287 & -0.272 & -0.318 & -0.283 & -0.166\\

$\alpha_{1}$ & -1.82 & -9.56 & 0.5 & 0.054 & 0.191 & 0.300 & 0.215 & 0.197 
& 0.209 & 0.182 & 0.208 & 0.208\\

$\alpha_{11}$ & -1.12 & -3.08 & 0.0 & -0.570 & -0.363 & -0.060 & -0.278 & 
-0.331 & -0.311 & -0.357 & -0.317 & -0.226\\

$\Phi$ & -0.47 & -0.5 & 0.5 & 0.376 & 0.389 & 0.641 & 0.643 & 0.449 & 0.450 
& 0.442 & 0.444 & 0.566\\

\end{tabular}
\end{center}
\end{table}


\begin{thebibliography}{10}
\item[$^*$] E-mail address: pylyp@icmp.lviv.ua
\item[$^\$$]E-mail address: huck@phys.sinica.edu.tw

\bibitem{notice1} There are a lot of publications dedicated to this theme. 
For brevity we do not present all of them here, but general review on 
critical behavior of infinite randomly dilute spin models is possible to 
find in \cite{PV00}.

\bibitem{PV00}
A. Pelissetto and E. Vicari, Phys. Rev. B {\bf 62},  6393  (2000).

 \bibitem{B83} K. Binder,  in {\em Phase Transitions and Critical 
Phenomena}, edited by C. Domb and J.~L. Lebowitz (Academic Press, London, 
1983), Vol.~8, pp.\ 1--144.

\bibitem{D86}
H.~W. Diehl,  in {\em Phase Transitions and Critical Phenomena}, edited by C.
  Domb and J.~L. Lebowitz (Academic Press, London, 1986), Vol.~10, pp.\
  75--267.

\bibitem{D97}
H.W.Diehl, Int.J.Mod.Phys.B {\bf 11}, 3503 (1997); preprint cond-mat/9610143.

\bibitem{notice2} The coupling $m_{0}$ is the "bare mass", representing a 
linear measure of the temperature difference from the critical point value.

\bibitem{DD81} H.W.Diehl and S.Dietrich, Z.Phys.B {\bf 42}, 65 (1981).

\bibitem{DD83} H.W.Diehl and S.Dietrich, Z.Phys.B {\bf 50}, 117 (1983).

\bibitem{DDE83} H.W.Diehl, S.Dietrich, and E.Eisenriegler, Phys. Rev. B 
{\bf 27}, 2937 (1983).

\bibitem{DC91} H.W.Diehl, A.Ciach, Phys.Rev.B {\bf 44}, 6642 (1991).

\bibitem{DSh98}
H.~W. Diehl and M. Shpot, Nucl. Phys. B {\bf 528},  595  (1998).

\bibitem{DN90} H. W. Diehl and A. N{\"u}sser, Z. Phys. B {\bf 79}, 69 
(1990); Z. Phys. B {\bf 79}, 79 (1990).

\bibitem{PS98} M. Pleimling and W. Selke, Eur. Phys. J. B {\bf 1}, 385 
(1998).

\bibitem{D98} H. W. Diehl, Eur.Phys.J. B {\bf 1}, 401 (1998).

\bibitem{ILSS98} F.Igl\'oi, P.Lajk\'o, W. Selke and F. Szalma, J. Phys. A 
{\bf 31}, 2801 (1998).

\bibitem{UShH00} Z.E. Usatenko, M.A. Shpot, Chin-Kun Hu, Phys. Rev. E {\bf 
63}, 056102 (2001).

\bibitem{UH02} Z.E. Usatenko and Chin-Kun Hu, Phys. Rev. E {\bf 
65}, 066103 (2002).

\bibitem{BL83} E.Brezin and S.Leibler, Phys.Rev.B {\bf 27}, 594 (1983).

\bibitem{RCz96} U.Ritschel and P.Czerner, Phys.Rev.Lett. {\bf 77}, 3645 
(1996).

\bibitem{CR97} A.Ciach and U.Ritschel, Nucl.Phys.B {\bf 489}, 653 (1997).

\bibitem{MCD99} A.Maciolek, A.Ciach, and A.Drzewinski, Phys.Rev.B {\bf 60}, 
2887 (1999).

\bibitem{Harris74}
A.~B. Harris, Journ. Phys. C {\bf 7},  1671  (1974).

\bibitem{B75} G.A.Baker, Jr., Essentials of Pad\'e Approximants (Academic, 
New York, 1975).

\bibitem{BNGM76} G.A.Baker, Jr., B.G. Nickel, M.S.Green, and D.I.Meiron, 
Phys. Rev. Lett. {\bf 36}, 1351 (1976).

\bibitem{L75}
T.~C. Lubensky, Phys. Rev. B {\bf 11},  3573  (1975).

\bibitem{GL76}
G. Grinstein and A. Luther, Phys. Rev. B {\bf 13},  1329  (1976).

\bibitem{BGZ76} E. Br\'ezin, J. C. Le Guillou, J.Zinn-Justin, Phase 
Transition and Critical Phenomena, ed.by C.Domb and M.S.Green, V.6 
(Academic Press, New York, 1976).

\bibitem{P80}
G. Parisi, J. Stat. Phys. {\bf 23},  49  (1980).

\bibitem{PS00} D.V.Pakhnin, A.I.Sokolov, Phys.Rev.B {\bf 61}, 15130 (2000).


\bibitem{DSh94} H.W.Diehl and M.Shpot, Phys. Rev. Lett. {\bf 73}, 3431 
(1994).

\bibitem{Sh97}
M. Shpot, Cond. Mat. Phys. N 10, 143  (1997).

\bibitem{Z89} J.Zinn-Justin, Euclidean Field Theory and Critical Phenomena 
(Oxford Univ. Press, New York, 1989).

\bibitem{Parisi88} G.Parisi, Statistical Field Theory (Addison-Wesley, 
Redwood City, 1988).

\bibitem{ID89} C.Itzykson and J.-M. Drouffe, Statistical Field Theory, Vol.I 
(Cambridge Univ. Press, Cambridge, 1989).

\bibitem{BB81} C.Bagnuls and C.Bervillier, Phys.Rev.B {\bf 24}, 1226 (1981).

\bibitem{mit1} The calculation of the surface correlation exponent 
$\eta_{\parallel}$ and other surface critical exponents of the special 
transition was performed in \cite{UH02}.

\bibitem{OO92}
K. Ohno and Y. Okabe, Phys. Rev. B {\bf 46},  5917  (1992).

\bibitem{Jug83}
G. Jug, Phys. Rev. B {\bf 27},  609  (1983).

\bibitem{mit2} The individual RG expansions for surface 
critical exponent was obtained from the above scaling relations 
 (\ref{28}) and (\ref{49}) (at $d=3$) and combining Eq.(\ref{48}) with the 
 $n\to 0$ limit of the two-loop series expansion for $\nu$
 $$\nu=\frac{1}{2}+\frac{n+2}{4(n+8)u}+\frac{v}{12}-
 \frac{1}{108}[\frac{(n+2)(38-27n)}{2(n+8)^2}u^2+\frac{11}{54}v^2+\frac{38-27n}{3(n+8)}uv].$$

\bibitem{BNM78} G.A.Baker, Jr., B.G. Nickel and D.I.Meiron, Phys. Rev. B 
{\bf 17}, 1365 (1978).

\bibitem{Sh88} N.A.Shpot, Phys.Lett.A {\bf 133}, 125 (1988); N.A.Shpot 
preprint ITF-88-16P (in Russian).


\end{thebibliography}
\end{document}